\documentclass[twocolumn]{aastex61}
\usepackage{color}

\usepackage{multirow}

\shortauthors{Sano, Yamane, Voisin et al.}

\begin{document}
\title{Discovery of Molecular and Atomic Clouds Associated with the Magellanic Superbubble 30~Doradus~C}


\author{H. Sano}
\affiliation{Institute for Advanced Research, Nagoya University, Furo-cho, Chikusa-ku, Nagoya 464-8601, Japan; sano@a.phys.nagoya-u.ac.jp}
\affiliation{Department of Physics, Nagoya University, Furo-cho, Chikusa-ku, Nagoya 464-8601, Japan; yamane@a.phys.nagoya-u.ac.jp}

\author{Y. Yamane}
\affiliation{Department of Physics, Nagoya University, Furo-cho, Chikusa-ku, Nagoya 464-8601, Japan; yamane@a.phys.nagoya-u.ac.jp}

\author{F. Voisin}
\affiliation{School of Physical Sciences, University of Adelaide, North Terrace, Adelaide, SA 5005, Australia; fabien.voisin@adelaide.edu.au}

\author{K. Fujii}
\affiliation{Department of Astronomy, School of Science, The University of Tokyo, 7-3-1 Hongo, Bunkyo-ku, Tokyo 133-0033, Japan}
\affiliation{National Astronomical Observatory of Japan, Mitaka, Tokyo 181-8588, Japan}

\author{S. Yoshiike}
\affiliation{Department of Physics, Nagoya University, Furo-cho, Chikusa-ku, Nagoya 464-8601, Japan; yamane@a.phys.nagoya-u.ac.jp}

\author{T. Inaba}
\affiliation{Department of Physics, Nagoya University, Furo-cho, Chikusa-ku, Nagoya 464-8601, Japan; yamane@a.phys.nagoya-u.ac.jp}

\author{K. Tsuge}
\affiliation{Department of Physics, Nagoya University, Furo-cho, Chikusa-ku, Nagoya 464-8601, Japan; yamane@a.phys.nagoya-u.ac.jp}

\author{Y. Babazaki}
\affiliation{Department of Physics, Nagoya University, Furo-cho, Chikusa-ku, Nagoya 464-8601, Japan; yamane@a.phys.nagoya-u.ac.jp}

\author{I. Mitsuishi}
\affiliation{Department of Physics, Nagoya University, Furo-cho, Chikusa-ku, Nagoya 464-8601, Japan; yamane@a.phys.nagoya-u.ac.jp}

\author{R. Yang}
\affiliation{Max-Planck-Institut f$\mathrm{\ddot{u}}$r Kernphysik, P.O. Box 103980, 69029 Heidelberg, Germany}

\author{F. Aharonian}
\affiliation{Max-Planck-Institut f$\mathrm{\ddot{u}}$r Kernphysik, P.O. Box 103980, 69029 Heidelberg, Germany}
\affiliation{Dublin Institute for Advanced Studies, 31 Fitzwilliam Place, Dublin 2, Ireland}

\author{G. Rowell}
\affiliation{School of Physical Sciences, University of Adelaide, North Terrace, Adelaide, SA 5005, Australia; fabien.voisin@adelaide.edu.au}

\author{M. D. Filipovi$\mathrm{\acute{c}}$}
\affiliation{Western Sydney University, Locked Bag 1797, Penrith South DC, NSW 1797, Australia}

\author{N. Mizuno}
\affiliation{National Astronomical Observatory of Japan, Mitaka, Tokyo 181-8588, Japan}

\author{K. Tachihara}
\affiliation{Department of Physics, Nagoya University, Furo-cho, Chikusa-ku, Nagoya 464-8601, Japan; yamane@a.phys.nagoya-u.ac.jp}

\author{A. Kawamura}
\affiliation{Department of Physics, Nagoya University, Furo-cho, Chikusa-ku, Nagoya 464-8601, Japan; yamane@a.phys.nagoya-u.ac.jp}

\author{T. Onishi}
\affiliation{Department of Astrophysics, Graduate School of Science, Osaka Prefecture University, 1-1 Gakuen-cho, Naka-ku, Sakai 599-8531, Japan}

\author{Y. Fukui}
\affiliation{Institute for Advanced Research, Nagoya University, Furo-cho, Chikusa-ku, Nagoya 464-8601, Japan; sano@a.phys.nagoya-u.ac.jp}
\affiliation{Department of Physics, Nagoya University, Furo-cho, Chikusa-ku, Nagoya 464-8601, Japan; yamane@a.phys.nagoya-u.ac.jp}

\begin{abstract}
We analyzed the 2.6-mm CO and 21-cm H{\sc i} lines toward the Magellanic superbubble 30~Doradus~C, in order to reveal the associated molecular and atomic gas. We uncovered five molecular clouds in a velocity range from 251 to 276 km s$^{-1}$ toward the western shell. The non-thermal X-rays are clearly enhanced around the molecular clouds on a pc scale, suggesting possible evidence for magnetic field amplification via shock-cloud interaction. The thermal X-rays are brighter in the eastern shell, where there are no dense molecular or atomic clouds, opposite to the western shell. The TeV $\gamma$-ray distribution may spatially match the total interstellar proton column density as well as the non-thermal X-rays. If the hadronic $\gamma$-ray is dominant, the total energy of the cosmic-ray protons is at least $\sim1.2 \times 10^{50}$ erg with the estimated mean interstellar proton density $\sim60$ cm$^{-3}$. In addition the $\gamma$-ray flux associated with the molecular cloud (e.g., MC3) could be detected and resolved by the Cherenkov Telescope Array (CTA). This should permit CTA to probe the diffusion of cosmic-rays into the associated dense ISM.
\end{abstract}


\section{Introduction} \label{sec:intro}
The Large Magellanic Cloud (LMC) is among the best laboratories for studying various astronomical objects and their physics because of its almost face-on inclination to us ($\sim 35^\circ$) and a well-known distance \citep[$\sim50$ kpc,][]{2013Natur.495...76P}. Note that many shell-like structures---supergiantshells, superbubbles, supernova remnants (SNRs), are located in the LMC and can be easily observed with very little contamination as compared to the Milky Way \citep[e.g.,][]{2001ApJS..136..119D,2013ApJ...763...56D,2016A&A...585A.162M,Bozzetto2017inprep}. Recent progress in multi-wavelength high angular resolution imaging has allowed us to study the relation between the various types of shells and their environments in the LMC.

We have already shown that the interaction between the shockwaves and the ambient interstellar gas, the so-called ``shock-cloud interaction,'' is an essential process for understanding the origins of high-energy radiation and cosmic-ray (CR) acceleration in Galactic SNRs. The shock-cloud interaction enhances the turbulence and magnetic field around gas clouds, producing rim-bright synchrotron X-rays and CR electrons with large roll-off energies \citep[e.g.,][]{2010ApJ...724...59S,2013ApJ...778...59S,2015ApJ...799..175S,2016arXiv160607745S}. Moreover, the surrounding interstellar gas becomes a target for CR protons to produce $\gamma$-rays via neutral pion decay \citep[][]{2012ApJ...746...82F,2013ASSP...34..249F,2013ApJ...768..179Y,2014ApJ...788...94F}. However, very few attempts have been made to study the Magellanic Clouds shells. As an important step, we are focusing on the Magellanic superbubble 30~Doradus~C and its surroundings to better understand the relation among the shock-cloud interaction, high-energy radiation, and the origin of CRs.

30~Doradus~C (DEM~L263, N157~C) is known as a bright X-ray shell in the LMC with a diameter of $\sim80$--100 pc \citep[e.g.,][]{2004ApJ...602..257B,2004ApJ...611..881S}. The non-thermal synchrotron X-ray shell is bright in its western region, whose luminosity is ten times higher than that of the Galactic SNR SN1006 \citep{2004ApJ...602..257B}. In contrast, the thermal X-ray shell is dominated by optically-thin thermal plasma and is bright in its eastern part, with estimated plasma ages of 0.9--6.7 $\times 10^5$ yr toward this region \citep{2004ApJ...602..257B,2004ApJ...611..881S,2009PASJ...61S.175Y}. The other bright X-ray sources 30~Doradus, N157B, and SN1987A are also located in the northeast and southwest of the superbubble.

30~Doradus~C also has been well-studied at multiple wavelengths. Optical and radio observations have resolved its shell-like morphology and revealed the presence of six stellar clusters, LH~90, with ages of 3--4 Myr and 7--8 Myr \citep[e.g.,][]{1985ApJS...58..197M,1993A&A...280..426T}. Most recently, H.E.S.S. detected $\gamma$-ray emission toward 30~Doradus~C with a 1--10-TeV luminosity of $(0.9\pm0.2) \times 10^{35}$ erg s$^{-1}$ \citep{2015Sci...347..406H}, suggesting evidence for CR proton and/or electron acceleration up to $\sim 100$ TeV. In contrast, the interstellar gas associated with 30~Doradus~C is yet to be studied in detail.

Here, we present $^{12}$CO($J$ = 1--0) results of 30~Doradus~C with the Mopra 22-m radio telescope. A detailed comparison of the CO, H{\sc i}, X-rays, and TeV $\gamma$-rays allows us to identify the associated molecular/atomic gas and to better understand the relationship among them.

\section{Observation} \label{sec:obs}
Observations of $^{12}$CO($J$ = 1--0) line emission at 115.271202 GHz were conducted from July 2014 to April 2015 using the Mopra 22-m radio telescope of the Australia Telescope National Facility. We used the OTF mode with Nyquist sampling, and the effective observation area was $7\arcmin \times 7\arcmin$. The typical system temperature was 700--800 K in the single-side band (SSB). The back end was the Mopra Spectrometer (MOPS) system with 4,096 channel (137.5-MHz) bands, corresponding to a velocity coverage of $\sim360$ km s$^{-1}$ and a velocity resolution of $\sim0.088$ km s$^{-1}$ ch$^{-1}$ in the zoom mode at 115 GHz. After convolution with a 2D Gaussian function, the final beam size was $\sim45\arcsec$ (FWHM). The pointing accuracy was checked every 2 h and was achieved to be within an offset of $\sim7\arcsec$. The absolute intensity was calibrated by observing Orion-KL [$\alpha_\mathrm{J2000}$ = $5^{\mathrm{h}}35^{\mathrm{m}}38\fs6$, $\delta_\mathrm{J2000}$ = $-5{^\circ}22\arcmin30\arcsec$] \citep{2005PASA...22...62L}. Finally, we combined the cube data with the {{MAGMA Data Release 3 (DR3) \citep[][]{2011ApJS..197...16W, Wong2017}}} using the root-mean-square weighting method. The final noise fluctuation was 0.022 K at a velocity resolution of 1 km s$^{-1}$.

\section{Results} \label{sec:res}
\begin{figure*}
\begin{center}
\includegraphics[width=\linewidth,clip]{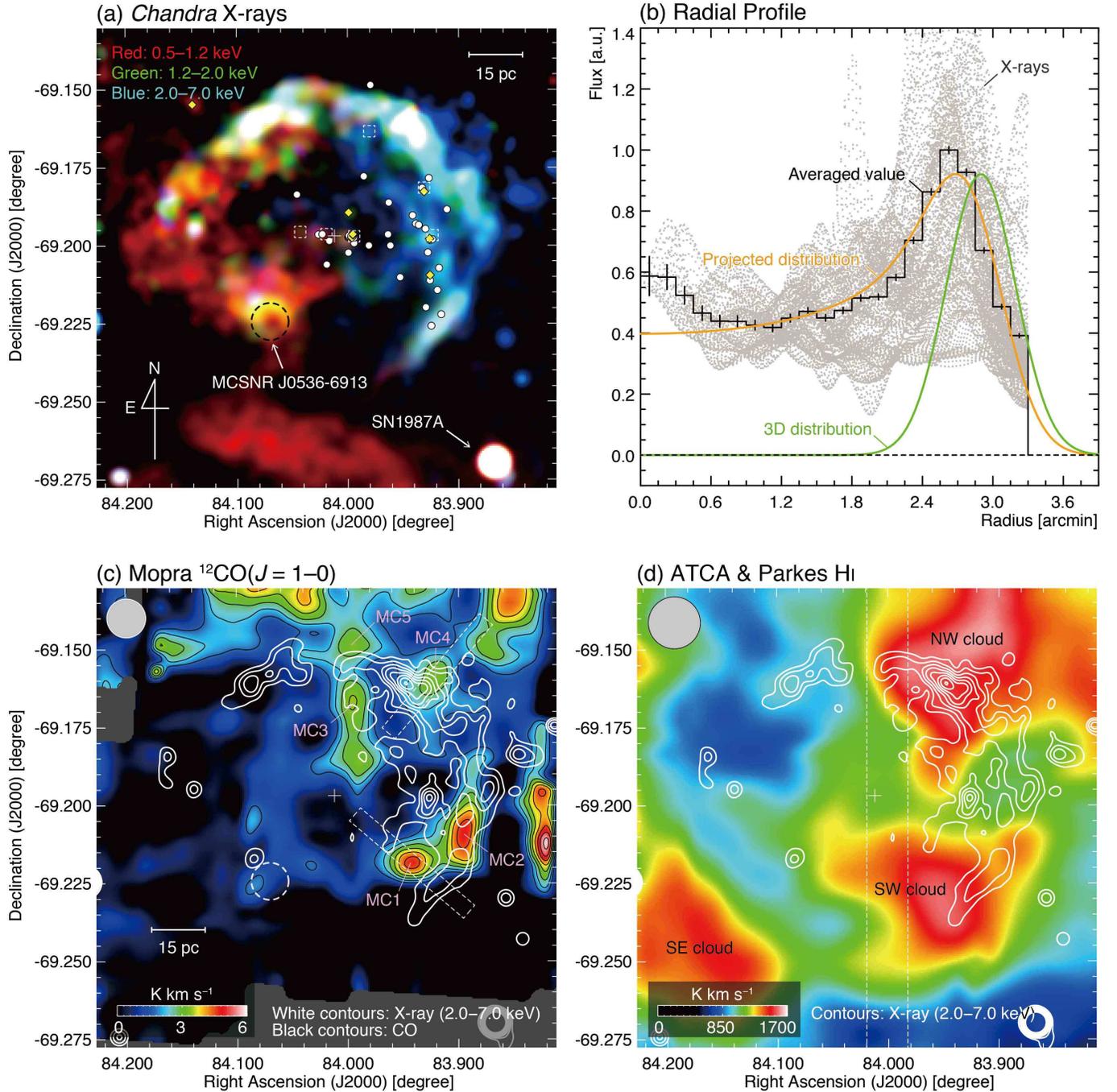}
\caption{(a) RGB image of the superbubble 30~Doradus~C observed by $Chandra$ \citep{2004ApJ...602..257B}. The red, green, and blue colors represent the energy bands, 0.5--1.2, 1.2--2.0, and 2.0--7.0 keV, respectively. The image is smoothed with a Gaussian function with a FWHM of 15$\arcsec$. The yellow diamonds, white filled circles, dashed squares, and dashed circles show the positions of Wolf-Rayet stars, O-type stars, LH~90, and MCSNR~J0536$-$6913, respectively. (b) Radial distribution of the broad-band X-rays (0.5--7.0 keV). Dotted plots show the distributions of all data points and solid lines show averaged values at each radius. Green and orange lines indicate the estimated 3D Gaussian distribution and its projected distribution, respectively. We calculated the radial distribution using only an azimuth angle clockwise from the east to the south. (c) Intensity distribution of $^{12}$CO($J$ = 1--0) obtained with Mopra overlaid with the $Chandra$ hard-band X-ray contours (black: 2.0--7.0 keV). The integration velocity range is from $V_\mathrm{LSR}$ = 251 to 276 km s$^{-1}$. The X-ray contours are from at $1.69 \times$ $10^{-8}$ counts s$^{-1}$ pixel$^{-1}$ and are square-root-spaced up to $13.2 \times$ $10^{-8}$ counts s$^{-1}$ pixel$^{-1}$. The white dashed line indicates regions for profile analysis in Figure \ref{radialprofile}. Black contours represent the CO integrated intensity. The lowest contour level and the contour interval are 2.80 K km s$^{-1}$ ($\sim 5 \sigma$) and 0.56 K km s$^{-1}$ ($\sim 1 \sigma$), respectively. (d) Intensity distribution of H{\sc i} obtained with ATCA $\&$ Parkes \citep{2003ApJS..148..473K}. The integration velocity range and overlaid contours are the same as in Figure \ref{four_images}(b). Crosses indicate the center of 30~Doradus~C.}
\label{four_images}
\end{center}
\end{figure*}%

Figure \ref{four_images}(a) shows an X-ray RGB image of 30~Doradus~C as observed by $Chandra$ \citep[e.g.,][]{2004ApJ...602..257B}. The soft-band X-rays (red; 0.5--1.2 keV, hereafter thermal X-ray) are dominated by the optically-thin thermal plasma ($\sim0.7$ keV), while the hard-band X-rays (blue; 2.0--7.0 keV) represent the non-thermal synchrotron X-rays from the CR electrons with energies of $\sim1$ TeV or higher \citep[e.g.,][]{2009PASJ...61S.175Y}. The thermal and non-thermal X-rays are bright in the eastern and western halves of 30~Doradus~C, respectively. These trends have also been investigated in the previous X-ray studies \citep{2004ApJ...602..257B,2004ApJ...611..881S,2009PASJ...61S.175Y,2015A&A...573A..73K}. In addition, six stellar clusters (LH~90), nine WR stars, 35 O-type stars, and their remnants (MCSNR~J0536$-$6913 and SN1987A) are part of the field of view shown Figure \ref{four_images}(a). The high-mass stars are mainly embedded in the western shell.

\begin{figure}
\begin{center}
\includegraphics[width=\linewidth,clip]{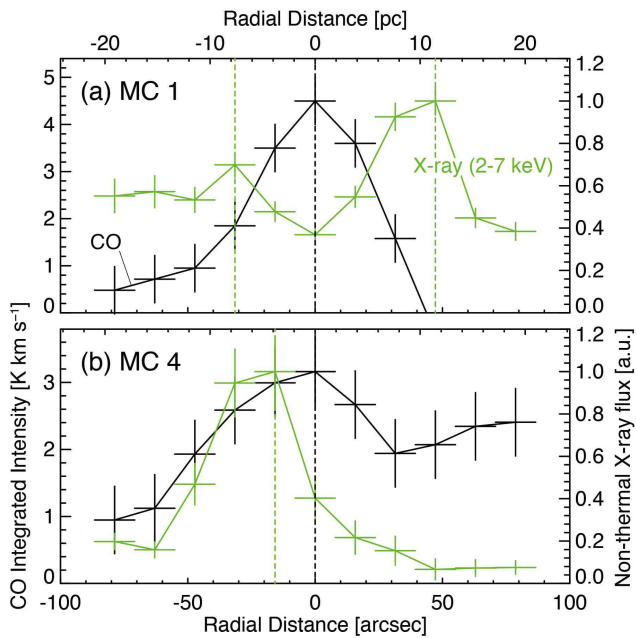}
\caption{Radial Profiles of the CO integrated intensity (black) and non-thermal X-ray flux in the energy band of 2.0--7.0 keV (green) for each rectangle region of (a) MC1 and (b) MC4, as shown by Figure \ref{four_images}c. Black and green dashed lines indicate the intensity peaks of CO and non-thermal X-rays, respectively.}
\label{radialprofile}
\end{center}
\end{figure}%

In order to estimate the radius and thickness (FWHM) of the X-ray shell, we assumed a three-dimensional spherical shell with a Gaussian distribution \citep[e.g.,][]{2012ApJ...746...82F}, and calculated only the region with an azimuth angle clockwise from the east to the south centered at ($\alpha_\mathrm{J2000}$, $\delta_\mathrm{J2000}$) = ($5^{\mathrm{h}}36^{\mathrm{m}}2\fs88$, $-69{^\circ}11\arcmin49\farcs2$), in which the X-rays with shell-like structure are significantly detected. Figure \ref{four_images}(b) shows the radial distribution of broad-band X-rays (0.5--7.0 keV). The actual structure of the shell named as ``3D distribution'' that we derived by Gaussian fitting the X-ray average value is measured from the center outwards. We found that the radius of $2\farcm89 \pm 0\farcm08$ ($\sim 42 \pm 1$ pc) and thickness of $0\farcm7 \pm 0\farcm2$ ($\sim 10  \pm 2$ pc) offer the best-fit values in Figure \ref{four_images}(b). Interestingly, the thickness of non-thermal dominant shell in the western half is 20 $\%$ less than that of thermal one in the eastern half.

Figure \ref{four_images}(c) shows the Mopra $^{12}$CO($J$ = 1--0) distribution in a velocity range from $V_{\mathrm{LSR}}$ = 251 to 276 km s$^{-1}$. Five CO clouds, MC1--5, were significantly detected ($>$ 5 $\sigma$) along with the western shell. Three of them are coincident with CO clouds named 30Dor-25, 26, and 28, which were reported in the CO survey of the 30~Doradus region \citep{1998J}. The physical properties of the molecular clouds are summarized in Table 1. We assumed a beam filling factor of 0.3 with $45''$ beam size \citep[e.g.,][]{2011AJ....141...73M,2015AA...580A..54O}.

\begin{figure*}
\begin{center}
\includegraphics[width=\linewidth,clip]{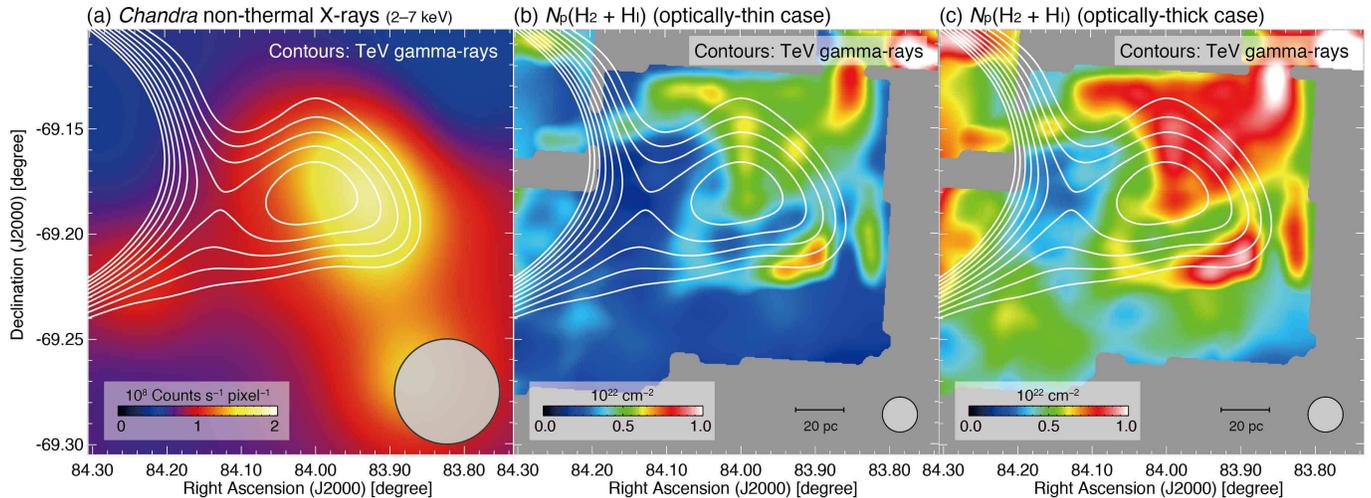}
\caption{(a) Distribution of $Chandra$ X-rays in the energy band of 2.0--7.0 keV overlaid on the H.E.S.S. TeV $\gamma$-ray contours in white \citep{2015Sci...347..406H}. The X-ray image was smoothed to match the point spread function of the TeV $\gamma$-rays ($\sim 180''$). Contour levels of TeV $\gamma$-rays are every 2 excess counts from 26 excess counts. (b) Distribution of total interstellar proton column density $N_\mathrm{p}$(H$_2$ + H{\sc i}) estimated from CO and H{\sc i} in a velocity range from $V_\mathrm{LSR}$ = 251 to 276 km s$^{-1}$. {{(c) The same image of Figure \ref{gamma_ism}(b), but considering the optically thick H{\sc i} (see the text).}} Overlaid contours {{of Figures \ref{gamma_ism}(b) and (c)}} are the same as Figure \ref{gamma_ism}(a).}
\label{gamma_ism}
\end{center}
\end{figure*}

It is remarkable that the CO distribution shows a good spatial correspondence with the non-thermal X-rays on a 10-pc scale. Furthermore, some of CO clouds are rim-brightened in the non-thermal X-ray region. Figure \ref{radialprofile} shows the radial profiles across the CO clouds MC1 and MC4. The profile x-axis increases towards the edge of the 30~Doradus~C shell. The origin position of the radial profile is defined as the maximum CO intensity in the projected distance. Negative and positive values represent to the inside and outside the shell, respectively. We can clearly see the non-thermal X-ray excess around the CO peaks. The physical separation between the CO and X-ray peaks are generally within 10 pc. This trend is not due to the interstellar absorption. The optical depth $\tau_{\mathrm{x}}$ for X-rays is expressed as follows \citep{1994hea2.book.....L};
\begin{eqnarray}
{\tau_{\mathrm{x}} = 2\times10^{-22}\phantom{0}N_{\mathrm{H}}\phantom{0}\mathrm{(cm^{-2})} \cdot \varepsilon^{-8/3} \phantom{0}\mathrm{(keV)},}
\end{eqnarray}

where $N_{\mathrm{H}}$ (cm$^{-2}$) is the interstellar proton column density, and $\varepsilon$ (keV) is photon energy of X-rays. $N_{\mathrm{H}}$ of the most intense molecular cloud MC1 is estimated to be $\sim 1.0 \times 10^{22}$ cm$^{-2}$ (see Figures \ref{gamma_ism}a and \ref{gamma_ism}b), which is consistent with the X-ray study by \cite{2015A&A...573A..73K}. The X-ray optical depths are calculated to be $\sim 0.3$ at 2 keV and $\sim 0.01$ at 7 keV, and hence the interstellar absorption effect is negligible.

\begin{deluxetable*}{lccccccccc}[]
\label{table}
\tablewidth{\linewidth}
\tablecaption{Properties of CO Clouds associated with 30~Doradus~C}
\tablehead{\multicolumn{1}{c}{Name} & $\alpha_{\mathrm{J2000}}$ & $\delta_{\mathrm{J2000}}$ & $T_{\rm R^\ast} $ & $V_{\mathrm{peak}}$ & $\Delta V$ & Size & Mass & $n$(H$_2$) & Comment\\
& ($^{\mathrm{h}}$ $^{\mathrm{m}}$ $^{\mathrm{s}}$) & ($^{\circ}$ $\arcmin$ $\arcsec$) & (K) & \scalebox{0.9}[1]{(km $\mathrm{s^{-1}}$)} & \scalebox{0.9}[1]{(km $\mathrm{s^{-1}}$)} & (pc) &  ($10^4$ $M_\sun $) & (cm$^{-3}$) &\\
\multicolumn{1}{c}{(1)} & (2) & (3) & (4) & (5) & (6) & (7) & (8) & (9) & (10)}
\startdata
\multirow{2}{*}{MC1} &  \multirow{2}{*}{05 35 42} & \multirow{2}{*}{$-69$ 13 8} & 0.37 & 250.4 & 6.0 & \multirow{2}{*}{6.2} & \multirow{2}{*}{{2.2}} & \multirow{2}{*}{{$\sim2600$}}  & \multirow{2}{*}{30Dor-28}\\
&   &  & 0.57 & 257.4 & 3.8 & & &\\
\multirow{2}{*}{MC2} &  \multirow{2}{*}{05 35 34} & \multirow{2}{*}{$-69$ 12 23} & 0.89 & 251.4 & 3.9 & \multirow{2}{*}{5.1} & \multirow{2}{*}{{1.5}} & \multirow{2}{*}{{$\sim3100$}}  & \multirow{2}{*}{30Dor-26}  \\
&   &  & 0.78 & 255.6 & 3.5 & & &\\
MC3 & 05 35 59 & $-69$ 10 23 & 0.38 & 273.1 & 4.9 &  {8.5} & {1.5} &  \phantom{0}$\sim700$ &\\
MC4 & 05 35 42 & $-69$ 09 38 & 0.53 & 268.6 & 3.1 & {8.7} & {1.4} &  \phantom{0}$\sim600$  &30Dor-25\\
MC5 & 05 35 59 & $-69$ 08 53 & 0.34 & 267.3 & 6.8 & 6.0 & {0.9} & {$\sim1200$} & \\
\enddata
\tablecomments{Col. (1): Cloud name. Cols. (2--7): Observed properties of the CO obtained by Gaussian fitting. Cols. (2)--(3): Position of the peak intensity. Col. (4): Maximum radiation temperature. Col. (5): Central velocity. Col. (6): FWHM line width. Col. (7): Cloud size defined as $(A / \pi)^{0.5} \times 2 \times F$, where $A$ is the total cloud surface area surrounded by the half intensity level of the maximum integrated intensity and $F$ is the beam filling factor of 0.3 (see text). Col. (8): The cloud mass is defined as $\mu$ $m_\mathrm{H} \sum_{i} [D^2$ $\Omega$ $N_\mathrm{i}(\mathrm{H_2})]$, where $\mu$ is the mean molecular weight, $m_\mathrm{H}$ is the mass of the atomic hydrogen, $D$ is the distance, $\Omega$ is the solid angle in a pixel, and $N$($\mathrm{H_2}$) is molecular hydrogen column density for each pixel. We used $\mu = 2.38$ by taking into account the helium abundance of {$36\%$} to the molecular hydrogen in mass, and $N$($\mathrm{H_2}$) = 7.0 $\times$ $10^{20}$[$W$($^{12}$CO) (K km $\mathrm{s^{-1}}$)] ($\mathrm{cm^{-2}}$) \citep{2008ApJS..178...56F}. (9) Number density of molecular hydrogen. (10) The names of the CO clouds 30Dor-25, 26, and 28 identified in \cite{1998J} are noted.}
\end{deluxetable*}

Figure \ref{four_images}(d) shows the H{\sc i} distribution obtained with ATCA $\&$ Parkes published by \cite{2003ApJS..148..473K}. We found three H{\sc i} clouds toward 30~Doradus~C in a velocity range of $V_{\mathrm{LSR}}$ = 251 to 276 km s$^{-1}$. Two of them are embedded within the northwest and southwest shells, while the southeast cloud seems to be distributed outside of the shell. The intensity peak of the northwest cloud shows good spatial correspondence with the non-thermal X-ray peak within the angular resolution of the H{\sc i} data. Note that the H{\sc i} intensity of the western shell is twice as high as that of the eastern shell. 

\begin{figure}
\begin{center}
\includegraphics[width=\linewidth,clip]{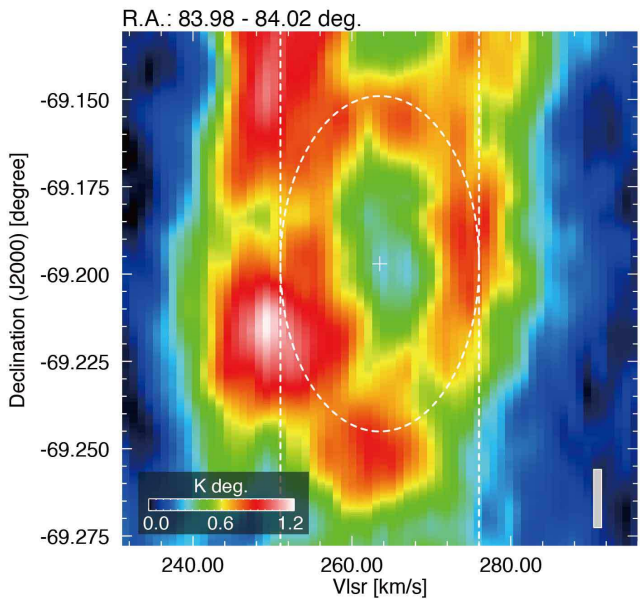}
\caption{Position-Velocity diagram of H{\sc i}. The integration range in the Right Ascension is from $83\fdg98$ to $84\fdg02$ shown
in Figure \ref{four_images}d. The dashed circle corresponds to the shell radius of 30~Doradus~C.}
\label{pvdiagram}
\end{center}
\end{figure}

Figure \ref{pvdiagram} shows the velocity-Declination diagram of H{\sc i}. We found an H{\sc i} cavity-like structure in the velocity range from 251 to 276 km s$^{-1}$, which has a similar diameter to 30~Doradus~C in terms of the Declination range. {{The expanding motion was also slightly seen in the CO clouds, but it is not highly significant due to a low signal to noise ratio of the current dataset.}}

Figure \ref{gamma_ism}(a) shows the $Chandra$ non-thermal X-rays (2.0--7.0 keV) superposed onto the H.E.S.S. TeV $\gamma$-ray contours in white \citep{2015Sci...347..406H}. The X-ray image was smoothed with a Gaussian kernel to match the point-spread function (PSF) of the H.E.S.S. data. The TeV $\gamma$-ray contours have two peaks; one is located in the center and the other is expanded to the left side in Figure \ref{gamma_ism}(a). The former is from 30~Doradus~C, and the latter is from the pulsar-wind nebula N157B. According to \cite{2015Sci...347..406H}, the TeV $\gamma$-ray excess toward 30~Doradus~C is considered to be independent from the $\gamma$-ray emission of N157B. For this study, we focused on the TeV $\gamma$-rays from 30~Doradus~C. In Figure \ref{gamma_ism}(a), it seems that the intensity peak of the TeV $\gamma$-rays is coincident with that of the non-thermal X-rays {within the large PSF of H.E.S.S.}. 

{Figures \ref{gamma_ism}(b) and (c)} show the distribution of total interstellar proton column density, $N_\mathrm{p}$(H$_2$ + H{\sc i}), obtained as $N_\mathrm{p}$(H$_2$ + H{\sc i}) = 2 $\times$ $N$(H$_2$) + $N_\mathrm{p}$(H{\sc i}), where $N$(H$_2$) and $N_\mathrm{p}$(H{\sc i}) are the number densities of molecular and atomic hydrogen, respectively. We used relation $N$(H$_2$) = 7.0 $\times$ 10$^{20}$ $\times$ $W$(CO), where $W$(CO) is velocity-integrated intensities of CO \citep{2008ApJS..178...56F}. On the other hand, the estimation of atomic hydrogen density would require more careful analysis. According to \cite{2015ApJ...798....6F}, 85$\%$  of H{\sc i} is optically thick ($\tau \sim 0.5$--3) in the Milky Way, and this trend also can be applied to the LMC. The averaged column density of H{\sc i} is 2--2.5 times higher than that derived by the optically thin assumption. Therefore, we used relationship $N_\mathrm{p}$(H{\sc i}) = $X$ $\times$ 1.823 $\times$ 10$^{18}$ $\times$ $W$(H{\sc i}), {where $X$ is scaling factor} and $W$(H{\sc i}) is the velocity-integrated intensities of H{\sc i}. {We estimated $N_\mathrm{p}$(H{\sc i}) in both cases of the optically thin ($X$ = 1, Figure \ref{gamma_ism}b) and the optically thick ($X$ = 2, Figure \ref{gamma_ism}c).} It can be clearly seen that $N_\mathrm{p}$(H$_2$ + H{\sc i}) is distributed from the center to the northwest of 30~Doradus~C, while there is much less gas in the southern and eastern regions. The $N_\mathrm{p}$(H$_2$ + H{\sc i}) distribution shows a spatial correspondence with the TeV $\gamma$-ray contours. The mean interstellar proton density is estimated to be {$\sim 40$ cm$^{-3}$ for the optically thin case; $\sim60$ cm$^{-3}$ for the optically thick case}, by assuming that the interstellar gas distribution also has a Gaussian shape similar to the X-rays with a radial extent of $\sim 47$ pc and a thickness of $\sim 10$ pc, where the radial extent is defined as the Gaussian peak radius ($\sim 42$ pc) $+$ HWHM of the Gaussian shell ($\sim 5$ pc). The fraction of molecular and atomic proton density is almost {$1 : 1$ for the optically thin case; $2 : 5$ for the optically thick case} over the whole 30~Doradus~C, while the density contribution of atomic proton is negligible toward the molecular clouds.

\section{Discussion}\label{sec:discussion}
\subsection{{Kinematics of the interstellar gas}}
{The CO/H{\sc i} cavity-like structure is generally considered to be a sign of expanding gas motion due to the stellar winds and supernova explosions. In fact, the cavity size is similar to that of 30~Doradus~C in declination (see the dashed circle in Figure \ref{pvdiagram}), indicating possible evidence for an expanding gas motion in 30~Doradus~C. The result lends strong support for the physical connection of the molecular and atomic gases with 30~Doradus~C.}

{The total mass of the interstellar gas associated with the superbubble is estimated to be $\sim2.2$--$3.4 \times 10^5$ $M_\odot$ within the radial extension of $\sim 47$ pc, the mass of atomic gas is $\sim1.2$--$2.4 \times 10^5$ $M_\odot$ and that of molecular gas is $\sim 1.3 \times 10^5$ $M_\odot$, where we assumed both optically thin/thick H{\sc i} and the helium abundance of $36\%$ in the molecular hydrogen mass \citep[c.f.,][]{2008ApJS..178...56F}. The momentum of total interstellar gas with the expansion velocity of 12.5 km s$^{-1}$ is calculated to be $\sim3.1$--$4.6 \times 10^{6}$ $M_\odot$ km s$^{-1}$. On the other hand, typical momentum of stellar winds from the O-type and Wolf-Rayet stars are $\sim 2.0 \times 10^{3}$ $M_\odot$ km s$^{-1}$ and $\sim 2.3 \times 10^{4}$ $M_\odot$ km s$^{-1}$, respectively \citep[c.f.,][]{1982ApJ...263..723A}. The stellar winds from the 11 Wolf-Rayet stars and 35 O-type stars within the shell \citep{1993A&A...280..426T} lead the total momentum of $\sim 3.2 \times 10^{5}$ $M_\odot$ km s$^{-1}$, which is only $\sim10\%$ of the momentum of the expanding interstellar gas.}

{We claim that the momentum missing can be interpreted as mainly arising from pre-existing gas motion. Most recently, \cite{2017arXiv170301075F} found that two H{\sc i}/CO velocity components known as the ``D-component'' and the ``L-component'' are colliding toward the eastern gas ridge of the LMC containing 30~Doradus~C. Typical velocity separation between the D- and L-components is from $\sim10$ to 50 km s$^{-1}$, which is consistent with the velocity separation of 30~Doradus~C ($\sim 25$ km s$^{-1}$). Moreover, the H{\sc i} clouds with $V_{\mathrm{LSR}} \sim 254$ km s$^{-1}$ and $\sim 274$ km s$^{-1}$ in the p-v diagram extend outside of 30~Doradus~C, particularly to the north of the superbubble; this cannot be predicted by the expanding H{\sc i} motion of 30~Doradus~C alone (see Figure \ref{pvdiagram}). We therefore conclude that the large-velocity separation was created by both the collision of the D- and L-components and the expanding gas motion of 30~Doraduc~C. It is therefore difficult to distinguish the expanding H{\sc i} motion from the pre-existing gas motion in this study alone. Further H{\sc i} and CO observations with a high angular resolution can reveal the kinematics of the interstellar gas associated with 30~Doradus~C in detail.}

\subsection{Origin of the non-thermal X-ray shell}
\label{subsec:physical}
The global properties of non-thermal X-rays have been well-described in the previous studies \citep{2004ApJ...602..257B,2004ApJ...611..881S,2009PASJ...61S.175Y,2015A&A...573A..73K}. The non-thermal X-rays are significantly detected over the whole superbubble, while the brightest non-thermal X-ray shell is located in the western region \citep[e.g.,][]{2015A&A...573A..73K}. The O-type and WR stars are embedded within the bright non-thermal shell, whose stellar mass function is extremely top-heavy with nine WR stars, 35 O-type stars, and 23 B-type stars \citep[c.f.,][]{1993A&A...280..426T}. Future, typical young Galactic SNRs have a non-thermal X-ray luminosity of $\sim 1 \times 10^{33}$ erg s$^{-1}$ \citep[e.g.,][]{2012ApJ...746..134N} ,while the non-thermal X-ray luminosity of 30~Doradus~C reaches $\sim1 \times 10^{34}$ erg s$^{-1}$ or higher \citep[e.g.,][]{2004ApJ...602..257B}. It is therefore likely that the non-thermal shell of 30~Doradus~C was created by multiple supernova remnants during the last few thousand years.

In addition, non-thermal X-ray peaks show good spatial correspondence with the molecular clouds MC1--5 on a 10-pc scale, while the CO peaks show spatial offsets from the non-thermal X-ray peaks on a 1-pc scale. These trends are possible evidence for the shock-cloud interaction that appear in the Galactic SNRs RX~J1713.7$-$3946 and RCW~86 \citep{2010ApJ...724...59S,2013ApJ...778...59S,2015ApJ...799..175S,2016arXiv160607745S}. In 30~Doradus~C, the post-shocked gas density was at most $n_0 \sim0.1$--0.4 cm$^{-3}$ as derived by the thermal X-rays in the southwest \citep{2004ApJ...602..257B,2015A&A...573A..73K}, whereas the interstellar proton density of the CO cloud MC2 is estimated to be $n_{\mathrm{cloud}} \sim 3100$ cm$^{-3}$. It is therefore suspected that the ambient gas inside the shell is completely evacuated by the strong stellar winds and several SNR shocks, while the CO clouds are survivors of shock erosion as the shock speed will stall toward the dense clouds as $(n_0/n_{\mathrm{cloud}})^{0.5}$ \citep[e.g.,][]{2010ApJ...724...59S}. The large velocity difference between the CO cloud surroundings and the inter-cloud space will enhance turbulence and magnetic field strength via the shock-cloud interaction. According to the three-dimensional magnetohydrodynamic simulations, the magnetic field strength will be amplified up to mG \citep{2009ApJ...695..825I,2012ApJ...744...71I}. The only difference between the Galactic SNRs and 30~Doradus~C is a spatial scale of X-ray enhancement. The physical separation of CO and X-ray peaks of 30~Doradus~C is ten times higher than that of Galactic SNR RX~J1713.7$-$3946 \citep[e.g.,][]{2010ApJ...724...59S,2013ApJ...778...59S}. One should consider the possibility that the separation was smoothed due to the low spatial resolution of CO dataset. Further radio observations using ALMA, ASTE, and ATCA will reveal the interstellar molecular and atomic gas distributions at $< 1$ pc resolution as well as the details of the shock-cloud interaction.

\subsection{Origin of the TeV $\gamma$-rays}\label{subsec:origing}
\begin{figure}
\begin{center}
\includegraphics[width=\linewidth,clip]{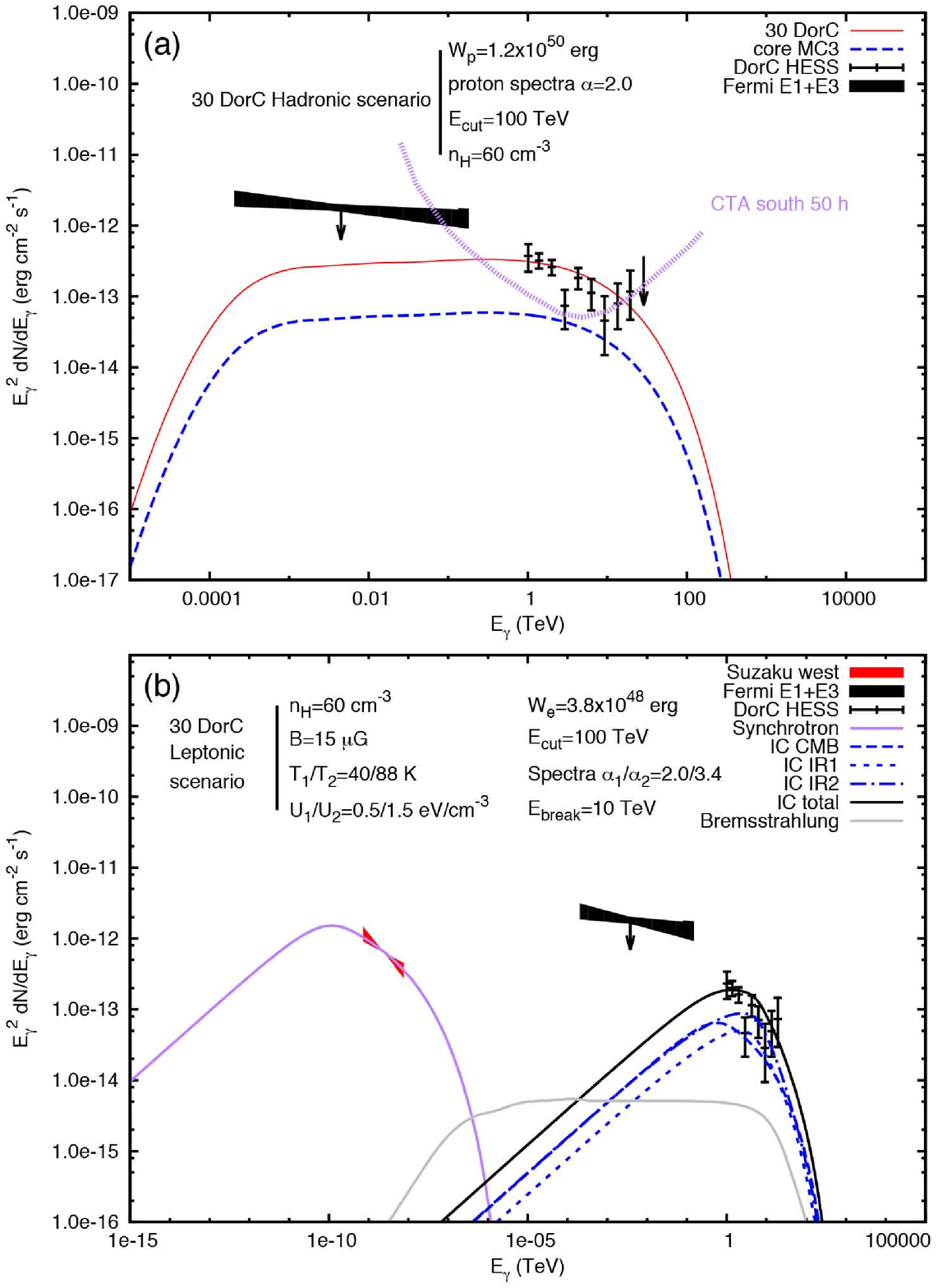}
\caption{SEDs of (a) hadronic scenario and (b) leptonic scenario used to models to fit the H.E.S.S. \citep[black points,][]{2015Sci...347..406H} and $Suzaku$ west \citep[red butterfly,][]{2009PASJ...61S.175Y}. Detailed parameters are shown in the text.}
\label{fig_sed}
\end{center}
\end{figure}

\cite{2015Sci...347..406H} recently announced TeV $\gamma$-ray detection towards 30~Doradus~C. Follow-up analysis with $Fermi$-LAT also resulted in the detection of an extended $\gamma$-ray source west of 30~Doradus~C \citep{2016A&A...586A..71A}. In this section, we combine our ISM results with the X-ray, GeV and TeV observations to discuss the plausible hadronic/leptonic nature of this TeV source. As we are however unsure about the association between the TeV and GeV sources, we thus use the $Fermi$-LAT spectral flux (see black butterfly in Figure \ref{fig_sed}) as a upper-limit.

In the case of a hadronic scenario, $\gamma$-ray emission originates from the decay of neutral pions produced by the CR-ISM inelastic interaction. Assuming an averaged ISM proton density $N_\mathrm{p}$(H$_2$ + H{\sc i}) = 60 cm$^{-3}${, optically thick case,} (see previous section) and a power-law distribution of CRs with spectral index $\alpha$ = 2.0, we note from Figure \ref{fig_sed}(a) that the hadronic SED reproduces the H.E.S.S. observation for a total energy budget $W_\mathrm{p}$ = $1.2 \times 10^{50}$ erg and a CR distribution energy cut-off $E_\mathrm{cut}$ = 100 TeV. The hadronic gamma-ray flux being sensitive to the ISM proton density, the upcoming  Cherenkov Telescope Array \citep[CTA,][]{2013APh....43....3A} with its arc-min spatial resolution and increased sensitivity could help highlight the $\gamma$-ray emission produced inside the various aforementioned molecular clouds. As an example we focus on the cloud MC3 as it spatially overlaps the TeV emission and is similar in angular scale to the CTA resolution. With CTA's ability to achieve arc-min scale angular resolution, it may be able to resolve the TeV $\gamma$-ray emission towards molecular cloud cores and hence probe the diffusion properties of cosmic-rays into such cores \citep[see discussion by][]{2007Ap&SS.309..365G,2012ApJ...744...71I}. Based on the volume ratio between the cloud MC3 and the superbubble 30~Doradus~C, we assume that a fraction $\sim5.0 \times 10^{-3}$ of the CR energy budget $W_\mathrm{p}$ resides inside this molecular cloud. Using the previous CR spectral shape, we observe that the flux produced inside molecular cloud MC3 (see dashed line in Figure \ref{fig_sed}(a)) could be detected, and perhaps resolved by CTA after at least 50 hours of observations.

Alternatively, $\gamma$-ray emission is also produced by high energy electrons from Bremsstrahlung and inverse-Compton radiation. As the type (continuous/impulsive) and age of the high energy source remains unconstrained, we intentionally do not use a time-dependent evolution of the energy distribution of electrons as  \cite{2015Sci...347..406H}. In order to account for the effect of radiative cooling at high energies (i.e. synchrotron and inverse-Compton), we instead use a broken power-law distribution $\mathrm{d}N_\mathrm{e}/\mathrm{d}E \propto (E/E_\mathrm{break})^{- \alpha_i} \mathrm{exp}(-E/E_\mathrm{cut})$ with $\alpha_i$ = $\alpha_1$ for $E < E_\mathrm{break}$ and $\alpha_i = \alpha_2$ for $E \geq E_\mathrm{break}$. As per \cite{2015Sci...347..406H} we also assume two populations of IR target photons coming from 30~Doradus~C ($T_1$ = 40 K, $U_1$ = 0.5 eV cm$^{-3}$) and the Tarantula nebula ($T_2$ = 88 K, $U_2$ = 1.5 eV cm$^{-3}$ ). As shown in Figure \ref{fig_sed}(b), the leptonic SED matches the X-ray and TeV data if we assume a distribution of electrons with energy budget $We = 3.8 \times 10^{48}$ erg, spectral indices $\alpha_1$ = 2.0 and $\alpha_2$ = 3.4, $E_\mathrm{break}$ = 10 TeV, $E_\mathrm{cut}$ = 100 TeV and an averaged magnetic field $B$ = 15 $\mu$G. We note that the Bremsstrahlung contribution is negligible at high energies. Finally, we argued in the previous section that the X-ray detections are non-thermal and can thus be explained by the propagation of high energy electrons towards the dense molecular clouds and their enhanced magnetic fields \citep[e.g.,][]{2010ApJ...725..466C}. Consequently, they are expected to undergo severe energy losses from synchrotron and IC radiation and, unlike the hadronic scenario, no inverse-Compton TeV emission should then be detected by CTA inside the molecular clouds.

\section{Conclusions} \label{sec:sum}
We summarize the present work as follows;

\begin{enumerate}
\item We identified molecular and atomic gas associated with the superbubble 30~Doradus~C using the Mopra $^{12}$CO ($J$ = 1--0) and ATCA $\&$ Parkes H{\sc i} datasets. Five CO clouds are distributed along the non-thermal X-ray shell in the west, while three of the H{\sc i} clouds are located at the northwest, southwest, and southeast.
\item The thermal X-rays are brighter in the eastern shell, where there are no dense CO/H{\sc i} clouds; conversely, the western shell has dense CO/H{\sc i} clouds and no evidence for thermal X-rays. The non-thermal X-rays are clearly enhanced around the molecular clouds on a pc scale, suggesting possible evidence for magnetic field amplification via the shock-cloud interaction.
\item The TeV $\gamma$-ray peak exhibits a good spatial correspondence with the distribution of the total interstellar proton column density as well as that of the non-thermal X-rays. If the hadronic process dominates, the total energy of the cosmic-ray protons will amount to at least $\sim1.2 \times 10^{50}$ erg with the mean interstellar proton density of $\sim60$ cm$^{-3}$.
\item The $\gamma$-ray flux associated with the molecular cloud (e.g., MC3) could be detected and resolved by CTA. This should permit CTA to probe the diffusion of cosmic-rays into the associated dense ISM \citep{2007Ap&SS.309..365G,2012ApJ...744...71I}.
\end{enumerate}

{\footnotesize{The Mopra radio telescope is part of the Australia Telescope National Facility. The University of New South Wales, the University of Adelaide, and the National Astronomical Observatory of Japan Chile Observatory supported operations. This work was financially supported by Grants-in-Aid for Scientific Research (KAKENHI) of the Japanese society for the Promotion of Science (JSPS, grant Nos. 15H05694 and 16K17664). This work also was supported by ``Building of Consortia for the Development of Human Resources in Science and Technology'' of Ministry of Education, Culture, Sports, Science and Technology (MEXT, grant No. 01-M1-0305).}}

\end{document}